\documentclass[12pt]{article}
\usepackage{graphicx}
\usepackage{natbib} 
\usepackage{url} 
\usepackage{amsmath, amssymb}

\DeclareMathOperator*{\argmin}{argmin}

\newcommand{\blind}{0}

\begin{document}

\def\spacingset#1{\renewcommand{\baselinestretch}%
{#1}\small\normalsize} \spacingset{1}


\if0\blind
{
  \title{Coupling material and mechanical design processes via computer model calibration}
  \author{Carl Ehrett\thanks{
    The authors gratefully acknowledge grant CMMI-1934438 from the National Science Foundation (NSF). CE was supported by fellowships through Department of Education GAANN grant P200A150310 and NSF NRT grant 1633608. DAB is also supported by NSF grants EEC-1744497 and OIA-1826715.}\hspace{.2cm}\\
    School of Mathematical and Statistical Sciences, Clemson University,\\
    D. Andrew Brown \\
    School of Mathematical and Statistical Sciences, Clemson University,\\
    Evan Chodora \\
    Department of Mechanical Engineering, Clemson University,\\
    Christopher Kitchens \\
    Department of Chemical and Biomolecular Engineering, Clemson University,\\
    and \\
    Sez Atamturktur \\
    Department of Architectural Engineering, Pennsylvania State University\\}
  \maketitle
} \fi

\if1\blind
{
  \bigskip
  \bigskip
  \bigskip
  \begin{center}
    {\LARGE\bf Coupling material and mechanical design processes via computer model calibration}
\end{center}
  \medskip
} \fi

\bigskip
\begin{abstract}
%
%
Computer model calibration typically operates by choosing parameter values in a computer model so that the model output faithfully predicts reality. 
	By using performance targets in place of observed data, we show that calibration techniques can be repurposed to wed engineering and material design, two processes that are traditionally carried out separately.
	This allows materials to be designed with specific engineering targets in mind while quantifying the associated sources of uncertainty. 
	We demonstrate our proposed approach by ``calibrating'' material design settings to performance targets for a wind turbine blade.
\end{abstract}

\noindent%
\vfill

\newpage
\spacingset{1} 
\section{Introduction}
\label{introduction}

Real-world optimization problems typically involve multiple objectives. 
This is particularly true in the design of engineering systems, where multiple performance outcomes are balanced against budgetary constraints. 
Among the complexities of optimizing over multiple objectives is the effect of uncertainties in the problem. 
Design is guided by models known to be imperfect, systems are built using materials with uncertainty regarding their properties, variations occur in the construction of designed systems, and so on. 
These imperfections, uncertainties and errors cause uncertainty also in the solution to a design problem. 
In traditional engineering design, one designs a system after choosing a material with appropriate properties for the project from a database of known materials. 
%
%
As a result, the design of the system is constrained by the initial material selection.
By coupling material discovery and engineering system design, we can combine these two traditionally separate processes under the umbrella of a unified multiple objective optimization problem.
In this paper, we cast the engineering design problem in the framework of computer model calibration.
In traditional calibration, one aligns computer model output to observations of a real system by estimating unknown parameters in the model.
Here, we instead align the computer model to performance and cost targets by finding design variables that optimize the model output with respect to those targets.
Our proposed methodology uses the Bayesian framework first established as a means for computer model calibration by Kennedy and O'Hagan \cite{Kennedy2001}.
This area is furthered by Higdon et al., \cite{Higdon2004}, who undertake model calibration with quantified uncertainty. 
%
%
Their approach is further refined and exemplified by Williams et al.\ \cite{Williams2006}.
Loeppky et al.\ \cite{Loeppky2006} offer a maximum-likelihood-based alternative to the Bayesian approach advocated by Kennedy and O'Hagan, intending thereby to improve the identifiability of the calibration parameters in the face of model discrepancy. 
Bayarri et al.\ \cite{Bayarri2007} extend the approach of Kennedy and O'Hagan, allowing for simultaneous validation and calibration of a computer model. 
Bayarri et al.\ \cite{Bayarri} apply this methodology to functional data using a hierarchical framework for the coefficients of a wavelet representation. 
Similarly, Paulo et al.\ \cite{Paulo2012} apply the approach of Ref.\ \cite{Bayarri2007} to computer models with multivariate output.
Brynsjarsd\'ottir et al.\ \cite{Brynjarsdottir2014} demonstrate the importance of strong priors on the model discrepancy term when undertaking calibration.
Common to those approaches is a conception of calibration as using real observations to get a posterior distribution on unknown parameters $\boldsymbol\theta$ so that the posterior predictive distribution of the model approximates reality.
By contrast, using an approach we call counterfactual Bayes, our methodology uses artificial observations (representing design targets) to obtain a posterior distribution on design variables $\boldsymbol\theta$ so that the posterior predictive distribution approaches those targets.
In counterfactual Bayes, we apply Bayesian reasoning to a hypothetical scenario that bears certain known relationships to reality.
Those known relationships allow us to transfer knowledge gained about the hypothetical scenario to reality, thereby gaining valuable insights into the phenomenon of interest.
We describe how, with little added computational cost, the methodology provides an initial rough estimate of the {\em Pareto front} for the system, i.e., the set of points in the design space that are Pareto optimal. 
A point is Pareto optimal if and only if, in order to improve any one of its elements, some other element must be made worse off.
%
%
In a system in which there is a trade-off between cost and performance, for example, which region of the Pareto front is most desirable will depend upon budgetary constraints.
This initial rough estimate of the Pareto front can be used to select artificial observations closer to the design space and thereby promote stronger Bayesian learning about the optimal settings for the design variables $\boldsymbol\theta$.
Repeated applications of the procedure can be used to produce the more thorough ``Pareto bands'' which estimate the Pareto front with quantified uncertainties.
%

Our approach is thus an example of Bayesian multi-objective optimization under uncertainty. Concerns about uncertainty in optimization may include uncertainty in the inputs (as when the inputs are not under perfect control), uncertainty in the outputs (as when the code or process of interest is not deterministic), and observation error \cite{Jin2005}. 
As a response to these sources of uncertainty, two different approaches are possible.
Firstly, one's focus may be on finding solutions that are robust to small perturbations in the inputs, e.g. by resampling in the design space around each candidate point considered during optimization \cite{Deb2006}.
Alternatively, rather than seeking robust solutions, one could endeavor simply to quantify the resulting uncertainty in the model outputs, again through resampling around points of interest \cite{Zhou2011}.
Our focus is on the latter of these two goals, though the posterior distributions our method provides include information about the sensitivity of the model output to the various model inputs, and thereby provides information about the robustness of the estimated optima.
Since we are concerned with computationally expensive objective functions, our method also avoids resampling, which can add prodigiously to the computational expense of an optimization procedure.
Though a Bayesian means of optimization, our proposed methodology contrasts with what is typically referred to as ``Bayesian optimization" (BO). 
In traditional BO, a Gaussian process (GP) surrogate model is constructed based on a small set of training observations, and the resulting updated GP is used to define an ``acquisition function'' that is used sequentially to select new observation locations until a stopping condition is achieved \cite{Picheny2019}.
Acquisition functions are crafted to attempt to balance exploration with exploitation of the objective function.
Examples include the EGO (efficient global optimization) function of Jones et al.\ \cite{Jones1998}, and the EGO-inspired SUR (stepwise uncertainty reduction) function showcased for univariate BO by Chevalier et al.\ \cite{Chevalier2014}.
The SUR approach is applied to estimating the Pareto front of a multi-objective optimization problem by Picheny \cite{Picheny2015}.
The methodology we propose here differs from these forms of traditional BO by its avoidance of sequential sampling, which is desirable in cases where the computational budget is very small, or when the relevant observations are based on previously gathered data, or in general when the data-gathering process is not devoted exclusively to serving the ends of the optimization.
Our methodology also can be used to quantify all associated forms of uncertainty discussed above -- uncertainty due to the model inputs, due to the stochastic nature of the objective function, or due to observation error of the outputs (e.g. if the observations are of the phenomenon of interest itself rather than of a computer model thereof).
Our approach thus has affinities with that of Olalotiti-Lawal and Datta-Gupta \cite{Olalotiti-Lawal2015}.
Both approaches use Markov chain Monte Carlo (MCMC, \cite{Gelfand1990}) to explore a posterior distribution on the design inputs.
Olalotiti-Lawal and Datta-Gupta construct a distribution that is designed to lie both on and near the Pareto front of the objective function.
The acceptance rate during MCMC and the variance of the distribution depend upon a user-defined temperature parameter.
The resulting posterior distribution includes uncertainty quantification.
The uncertainty quantified in that approach is the uncertainty remaining in the distribution designed by the authors; this distribution does not itself come from the model of the phenomenon of interest.
By contrast, under our approach, the distribution explored via MCMC is dictated by the model itself (and by the GP surrogate thereof), by our prior knowledge about the appropriate design settings, and by the choice of performance/cost targets.
The approach we propose here also has somewhat greater flexibility -- while it may be used to explore the entire Pareto front as Olalotiti-Lawal and Datta-Gupta do, our approach also may be used as a form of ``goal programming" \cite{Miettinen2008}, targeting a particular region of the Pareto front in accordance with the preferences of the relevant decision-makers.
%

%
We apply our proposed methodology both to a proof-of-concept example and to finding material design settings to optimize performance and cost for a wind turbine blade of fixed outer geometry.
The blade is to be constructed using a composite material, the properties of which are dependent upon design variables under our control.
%
%
Our material design goal is to reduce the cost per square meter of the composite, the angle of twist (in radians) of the blade when under load, and the deflection (in meters) of the blade tip when under load.
In Section \ref{counterfactual_bayes}, we introduce the counterfactual Bayes methodology for learning about a real system by applying Bayesian reasoning in a hypothetical scenario with known linkages to the real system.
In Section \ref{calib_for_design}, we review the calibration framework grounding our design optimization approach. 
In Sections \ref{example} and \ref{application}, we apply our methodology to simulated data and to wind turbine blade design.
%
%
Section \ref{conclusion} discusses the results and thoughts about future directions.
\section{Counterfactual Bayes}\label{counterfactual_bayes}
Counterfactual Bayes relies on reasoning about counterfactual situations.
In this respect it is reminiscent of some statistical approaches, such as the framework of Rubin \cite{Rubin1974}, to questions of causality.
To elucidate the relevant notion of counterfactuality, 
we rely on the logician's conception of possible worlds, which are used to explain the semantics of modal sentences (sentences having to do with possibility and necessity, rather than what is merely actual). 
A possible world can be conceived of as an internally consistent description of a world, which may or may not match the actual world \cite{Adams1974,Lewis1986}. 
So even though it is perhaps true that all dogs weigh under 350lbs (the heaviest recorded dog weighed 343lbs \cite{Young1994}), there is a possible world in which some dogs weigh over 350lbs, since it is possible to describe such a world without contradicting oneself.
I.e., a 350lb dog \textit{could} exist.
%
%
By contrast, there is no possible world in which some dogs are reptiles. 
Dogs are mammals by definition, and to describe any creature simultaneously as a reptile and as a dog is to contradict oneself. 
%
%
%
%

%
Possible worlds give substance to the fundamental concepts of counterfactual approaches to causality.
For example, consider a case in which event $A$ occurs at time $t$, event $B$ occurs at time $t+1$, and one hypothesizes that $A$ caused $B$.
This causal claim, in the language of possible worlds, is equivalent to something like the following: In any possible world $\omega$ which is identical to our world $\alpha$ at all points up to time $t$, and which differs at time $t$ only in that $A$ did not occur, is also such that $B$ does not occur in $\omega$.
The causal claim is thus implicitly a claim about worlds other than our own, based on observations made in our own world.
Counterfactual Bayes, as described in this paper, essentially reverses this relationship.
The idea of counterfactual Bayes is to rely on observations made in other possible worlds in order to learn about features of our own world.
We can summarize the methodology of counterfactual Bayes as follows.
Let $\alpha$ denote the actual world. 
Let $f_\alpha$ be a function relating inputs $\mathbf x,\boldsymbol \theta$ to some output $\mathbf y$, which describes some phenomenon of interest in the real world for which we wish to find optimal settings for $\boldsymbol\theta$. 
Suppose that $f_\alpha$ is such that the optimal outcome can be redescribed in terms of some target outcome $\mathbf y_t$, i.e., that $\argmin_{\boldsymbol\theta} f_\alpha(\mathbf x,\boldsymbol \theta)=\argmin_{\boldsymbol\theta} \lVert \mathbf y_{t} - f_\alpha(\mathbf x,\boldsymbol\theta)\rVert$. 
Then a distribution $\boldsymbol\theta|\mathbf x,\mathbf y_{t}$ \textit{de facto} approximates a distribution on $\boldsymbol\theta$ values producing the optimal achievable output of the system. 
%
%
Consider now a possible world $\omega$ in which the same phenomenon $f_\omega=f_\alpha$ holds true, and in which we observe $\mathbf y_{t}$. 
Then we can apply Bayes' rule to learn a posterior distribution $p(\boldsymbol\theta|\mathbf x,\mathbf y_{t})$ of $\boldsymbol \theta$ values in $\omega$. 
This is not directly applicable to the actual world, since we have not observed $\mathbf y_{t}$ here. 
But we have that $\boldsymbol\theta|\mathbf x,\mathbf y_{t}$ approximates a distribution on $\boldsymbol \theta$ values producing an optimal achievable outcome from the system $f_\omega$, and we furthermore have that $f_\alpha=f_\omega$ and thus that a distribution on $\boldsymbol \theta$ values optimal for $f_\omega$ is also a distribution on $\boldsymbol \theta$ values optimal for $f_\alpha$. 
Thus by relying on known connections between $\omega$ and $\alpha$, we use observations made only in $\omega$ to gain valuable insight into features of $\alpha$.
In what follows, we apply this counterfactual Bayes approach to find distributions on optimal design settings.
In our approach, Kennedy-O'Hagan-style model calibration \cite{Kennedy2001} is what allows us to discover a posterior distribution on design settings $\boldsymbol\theta$ given an ``observation'' of a set of possible outcomes.
We will apply that model calibration framework in a hypothetical scenario involving artificial observations of idealized outcomes $\mathbf y_t$, using our knowledge of the true system to exploit the resulting posterior distribution $\boldsymbol\theta|\mathbf y_t$ and thereby discover a distribution on optimal design settings.
\section{Calibration for design}\label{calib_for_design}

%
\subsection{Gaussian process emulators for calibration}
In this work, when an emulator is needed we use GP emulators.
As a multivariate Gaussian random variable is characterized by a mean vector and covariance matrix, a GP is characterized by a mean and covariance functions $\mu:\mathcal D\to \mathbb R$ and $C:\mathcal D\times \mathcal D\to \mathbb R$, where $\mathcal D$ is the domain of the process. 
For points $\mathbf x,\mathbf y\in \mathcal D$, $\mu(\mathbf x)$ is the GP mean at $\mathbf x$, and $C(\mathbf x, \mathbf y)$ is the covariance between the values of the GP at $\mathbf x$ and $\mathbf y$.
The distribution of the GP at any finite number of points is multivariate normal with mean vector and covariance matrix determined by $\mu(\cdot)$ and $C(\cdot,\cdot)$.
In principle, model calibration need not rely on emulators; one can complete a Bayesian analysis via MCMC by running the model at each iteration of the chain (see e.g. Ref.\  \cite{Hemez2011}). 
%
In Section \ref{example} we assume fast-running computer code for the simulated example, but
computer models are often too computationally expensive to allow such expenditure \cite{VanBuren2013,VanBuren2014}.
Instead, a computationally tractable emulator can be trained using a sample of the computer model output. 
GPs are popular prior distributions on computer model output for three reasons.
Firstly, their use does not require detailed foreknowledge of the model function's parametric form. 
Secondly, GPs easily interpolate the computer model output, which is attractive when the model is deterministic and hence free of measurement error. 
This is the usual case, although some attention has focused on calibrating stochastic computer models \cite{Pratola2018}. 
Thirdly, GPs facilitate uncertainty quantification through the variance of the posterior GP. 
This section provides brief background on GPs and their use in regression broadly, and in computer model calibration specifically.
The use of GPs as a computationally efficient predictor of computer code given observations of code output is advocated by Sacks et al.\ \cite{Sacks1989} and explored at length by Santner et al.\ \cite{Santner2003a}.
Since computer code is typically deterministic, these applications differ from the focus of Ref.\ \cite{OHagan1978} in that the updated GP interpolates the computer output. 
Ref.\ \cite{Kennedy2001} uses GPs for computer model calibration. 
Kennedy et al.\ \cite{Kennedy2006} showcase this use of GP emulators for uncertainty and sensitivity analyses. 
Bastos and O'Hagan \cite{Bastos2009} describe numerical and graphical diagnostic techniques for assessing when a GP emulator is successful, as well as likely causes of poor diagnostic results. 
Though most work on GP emulation uses stationary covariance functions 
and quantitative inputs, 
%
Gramacy and Lee \cite{Gramacy2008} use treed partitioning for a nonstationary computer model, and
Qian et al.\ \cite{Qian2008} explore methods that include both quantitative and qualitative inputs.
Whether or not an emulator is used, one may consider a computer model to be of the form $\eta(\mathbf x,\boldsymbol \theta)$, where $(\mathbf x,\boldsymbol \theta)$ comprise all model inputs. 
The vectors $\boldsymbol \theta$ and $\mathbf x$ denote respectively the inputs to be calibrated and the \emph{control inputs}, which are all other model inputs that are known and/or under the researchers' control. 
%
%
%
Thus, the model is
\begin{equation} \label{eq:model_gen}
y(\mathbf x)=f(\mathbf x)+\epsilon(\mathbf x) = \eta(\mathbf x,\boldsymbol \theta) + \delta(\mathbf x)+\epsilon(\mathbf x),
\end{equation} 
where $y(\mathbf x)$ is the observed response at control inputs $\mathbf x$, $f(\cdot)$ is the true system, $\delta(\cdot)$ is the model discrepancy (the systematic bias of the model) and $\epsilon(\cdot)$ is mean-zero error, often assumed to be i.i.d.\ Gaussian. 
To use an emulator, suppose we have inputs $\{(\mathbf x_i,\mathbf t_i)\}_{i=1}^n\subseteq \mathbb R^p\times \mathbb R^q$ scaled to the 
unit hypercube and completed model runs 
$\eta\left(\mathbf x_i,\mathbf t_i\right)$ for $i=1,\ldots,n.$
Define the GP prior for $\eta(\cdot,\cdot)$ as having mean function $\mu(\mathbf x,\mathbf t)=c$, where $c$ is a constant, and
set the covariance function in terms of the marginal precision $\lambda_\eta$ and a product power exponential correlation:
\begin{equation}\label{eq:Hig_cov}
\begin{split}
C((\mathbf x,\mathbf t),(\mathbf x',\mathbf t')) = &\frac 1\lambda_\eta \prod_{k=1}^{p}
\exp \left(-\beta^\eta_k|x_k-x_k'|^{\zeta_\eta}\right) \times\\
& \prod_{j=1}^{q}
\exp \left(-\beta^\eta_{p+j}|t_j-t_j'|^{\zeta_\eta}\right) +\\
&\sigma^2 \mathbf I_{(\mathbf x,\mathbf t)=(\mathbf x',\mathbf t')}
\end{split}
\end{equation}
where each $\beta_k$ describes the strength of the GP's dependence on one of the elements of the input vectors $\mathbf x$ and $\mathbf t$, and $\zeta_\eta$ determines the smoothness of the GP. 
Independent Gaussian observation error is captured by $\sigma^2$ and the indicator $\mathbf I$.
If $\eta(\cdot,\cdot)$ is a deterministic computer model, then $\sigma^2=0$.
The model is completed by specifying priors for the hyperparameters $c,\lambda_\eta,\alpha_\eta,\beta^\eta_j$ and $\sigma^2$ for $j=1,\ldots,p+q$, though in practice these are often set to predetermined values.
\subsection{Design to target outcomes}
Call design targets treated as observations in the design procedure we propose below ``target outcomes'', and call that procedure, which pairs a Bayesian model calibration framework with target outcomes via counterfactual Bayes, ``calibration to target outcomes" (CTO). 
Thus target outcomes are a sort of artificial data, and the calibration procedure is carried out as if these artificial data had been observed in reality.
As in traditional calibration, in which the result is a distribution on the calibrated parameter $\boldsymbol\theta$ to approximate the observed data, in CTO the result is a distribution on the design parameter $\boldsymbol\theta$ which induces the model to approximate the performance and cost targets.
%
%
%
%
%

%
%
%
%
%

The tools of model calibration as based on the work of Ref.\ \cite{Kennedy2001} retain their advantages under our proposed methodology.
Most centrally, calibrating to target outcomes $\mathbf y$ produces not merely a point estimate $\mathbf t^*$, but rather a posterior distribution of $\mathbf t|\mathbf y$ reflective of remaining uncertainty about the optimal value of $\mathbf t^*$. 
Such uncertainty may come from parameter uncertainty (uncertainty about the values of model inputs other than the design variables), model form uncertainty (uncertainty about how closely the code approximates reality), and observation error. 
%
%
%
The Bayesian model calibration framework allows for quantification of all of these uncertainties. 
In the Kennedy-O'Hagan framework, the goal is computer model calibration, so that $\eta(\cdot,\cdot)$ is a computer model representing some real phenomenon $f(\cdot)$. 
The framework is naturally suited to computer model calibration because $\boldsymbol\theta$ is an input for $\eta(\cdot,\cdot)$ but not for the real system of interest $f(\cdot)$.
By contrast, in CTO, $\boldsymbol\theta$ is an input for the real system of interest, since $\boldsymbol\theta$ is a design setting for the system.
Thus under CTO we may take $\eta(\cdot,\cdot)$ either to be a computer model as under KOH, or, alternatively, we may take $\eta(\cdot,\cdot)$ itself to be the real system of interest.
In either case, a set $\boldsymbol\eta$ of observations of $\eta(\cdot,\cdot)$ can be used to produce a GP model.
When $\eta(\cdot,\cdot)$ is the real system, we would have $\delta(\cdot)\equiv0$.
If $\eta(\cdot,\cdot)$ is a computer model, the process of calibrating that model takes place separately from CTO, and so for the purposes of CTO $\delta(\cdot)$ can be treated as known.
Equivalently, we can absorb the known discrepancy into the $\eta(\cdot,\cdot)$ term, again setting $\delta(\cdot)\equiv0$.
As a result, CTO is not afflicted by the identifiability concerns of the Kennedy-O'Hagan framework \cite{Bayarri2007,Tuo2016}.
It is common to plug in the MLEs of the GP covariance hyperparameters $\lambda_\eta$ and $\boldsymbol \beta^\eta$ in \eqref{eq:Hig_cov} instead of including them in a full Bayesian analysis \cite{Kennedy2001,Santner2003a,Qian2008,Paulo2012}.
In our proposed methodology, that is not merely a convenience, but rather is essential 
%
%
%
%
to avoid training an emulator using the target outcomes, which by their nature are extreme outliers (see Ref.\ \cite{Liu2009} on the dangers that arise here).
We use values found by maximizing the log likelihood of the available simulation runs with respect to $\lambda_\eta$ and $\boldsymbol\beta^\eta$.
We set the GP to have constant mean $c=0$, which works well when (as here) responses are standardized with mean 0, and the GP is not used for extrapolation \cite{Bayarri2007}.
We set $\zeta_\eta = 2$, which assumes that the model output is infinitely differentiable.
%


%
Denote completed runs of the simulator $\boldsymbol \eta = (\eta(\mathbf x_1,\mathbf t_1),\cdots,\eta(\mathbf x_n,\mathbf t_n))^T$, target outcomes $\mathbf y_t = (y(\mathbf x_{n+1}),\cdots,y(\mathbf x_{n+m}))^T$, 
and $D = (\boldsymbol \eta^T,\mathbf y_t^T)^T$.
Then $D | \boldsymbol \theta,\boldsymbol\sigma^2,\widehat{\lambda_\eta}, \widehat{\boldsymbol \rho^\eta}$ is multivariate normal with mean 0 and covariance $\mathbf C_D$, a matrix with $i,j$ entry equal to 
$
C((\mathbf x_i,\mathbf t_i),(\mathbf x_j,\mathbf t_j)) + 
\sigma^2 \mathbf I_{i=j>n}
$
%
$\mathbf I(\cdot)$ is the indicator function. 
Gaussian error variance $\sigma^2$ reflects our assumption that in $\omega$, the world in which the targets are observed, they are subject to observation error.
If $\boldsymbol\eta$ is subject to observation error, so that $C(\cdot,\cdot)$ itself includes nonzero observation error variance, then allowing $\sigma^2>0$ is not strictly necessary.
However, given that the targets may be extreme outliers, it is often computationally beneficial to include $\sigma^2$, thereby supposing that the target outcomes were observed with greater than usual observation error.
Whether this assumption is appropriate in a particular application will depend on the phenomenon of interest.
However, it is necessary to construct the model in such a way that the target outcomes are compossible with the model, and including $\sigma^2$ ensures that this requirement is satisfied even when $\boldsymbol\eta$ does not include observation error.
Whether or not $\sigma^2$ is included, it may be beneficial to include a small nugget into the covariance function $C(\cdot,\cdot)$.
This improves the conditioning of the covariance matrix. 
Where $\eta(\cdot,\cdot)$ has $m>1$ outputs, in many cases it suffices to fit a separate GP to each output.
We take this approach here, letting $\sigma^2_i$ be the observation error variance for the $i^\text{th}$ output.
Whether this assumption of independence among the outputs is appropriate will depend on the application.
We set a Gamma $(4,\mathrm{scale }=1/8)$ prior on each element of $\boldsymbol\sigma^2=(\sigma^2_1,\cdots,\sigma^2_m)^T$, encouraging low values of observation error variance, since high values would allow non-optimal regions of the design space to enjoy likelihood near to that of the optimal region.
Setting a uniform prior on the design variables $\boldsymbol\theta$, the joint posterior density under the model is
\begin{equation} \label{eq:full_dist}
\pi(\boldsymbol \theta,\boldsymbol \sigma^2| D,\widehat{\lambda_\eta},\widehat{\boldsymbol \rho^\eta})
\propto \pi(D | \boldsymbol \theta,\widehat{\lambda_\eta}, \widehat{\boldsymbol \rho^\eta}) \times \pi(\boldsymbol \sigma^2).
\end{equation}
MCMC methods are used to explore the posterior distribution.
When one has little information about the location and shape of the system's Pareto front in a multiobjective system, it may not be obvious what target best accords with one's goals.
One common choice in such situations is to locate the portion of the Pareto front closest to the ``utopia point''---the global minimum of each objective function.
When one has access to a set of observations $\boldsymbol\eta$, the utopia point can be estimated by taking the minima of the observations of each objective.
However, another option in such cases is to perform a ``preliminary round'' of CTO to estimate the system's Pareto front.
In preliminary CTO, one performs the usual CTO routine with a target known to be less than or equal to the utopia point of the system in all objectives, and with $\boldsymbol\sigma^2$ set to a large constant for each objective---e.g. approximating a flat prior on observation error by setting $\sigma^2_i =5\cdot10^7$ for $i=1,\ldots,m$ for $m$ objectives standardized to each have standard deviation 1 (under the uniform prior on $\boldsymbol\theta$).
This encourages CTO to explore broad regions of the feasible design space near the Pareto front.
When the resulting posterior samples of $\boldsymbol\theta$ are filtered to retain only their Pareto dominant subset, this forms a rough estimate of the Pareto front that can be used to select target outcomes in an informed way.
In addition to being only a rough estimate of the Pareto front, this preliminary estimate does not include quantification of uncertainties regarding its location.
Methods for estimating the system's Pareto front accurately with quantified uncertainties are explored in Section \ref{removing_cal_pars}.
The full CTO process, including preliminary Pareto front estimation, is given in Algorithm 1.

\begin{figure}[h]
	\centering
	\begin{tabular}{|p{.025\linewidth}|p{.85\linewidth}|}
		\hline
		\multicolumn{2}{|p{.9\linewidth}|}{Algorithm 1: Full CTO procedure including preliminary estimation of Pareto front}\\
		\hline
		1.& Set target outcomes $\mathbf y_t$ out of the feasible design space and $\sigma^2 = s\cdot(1, 1, \ldots, 1)$ for large constant $s$.\\
		2.&  Use MCMC to sample $\boldsymbol\theta|\mathbf y_t$ and thereby the posterior predictive distribution.\\
		3.&  Filter the predictions to retain only their Pareto optimal values $\mathcal P$. \\
		4.& Select new target outcomes $\mathbf y_t^*$ using $\mathcal P$ as an estimate of the model's Pareto front. \\
		5.& Setting $\sigma^2_i \sim \mathrm{gamma}(4,1/8)$ for $i=1,\ldots,m$, use MCMC to draw from $\boldsymbol\theta|\mathbf y_t^*$.\\
		\hline
	\end{tabular}
	\label{alg:CDO_alg}
\end{figure}

\begin{figure}
	\centering
	\includegraphics[scale=.85]{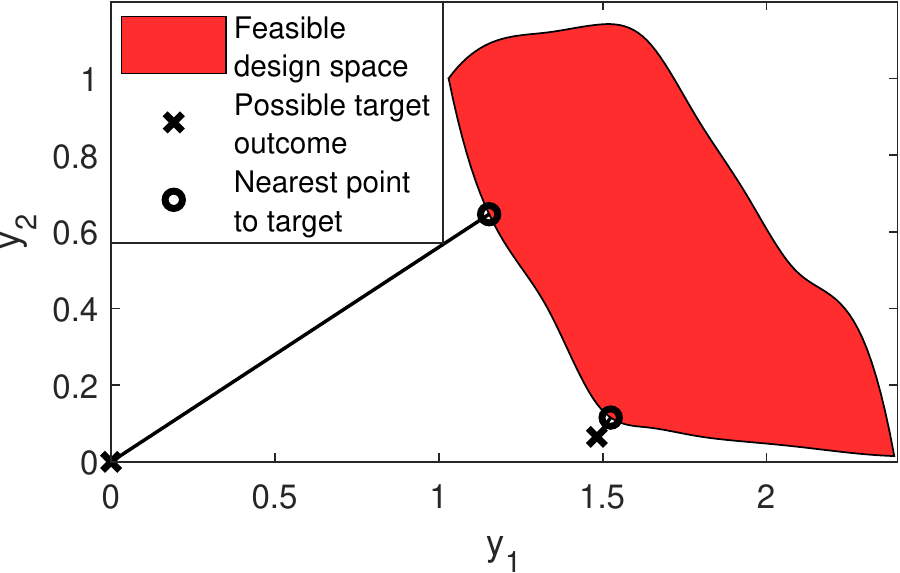}
	\caption{Two choices of target outcomes, drawing the posterior predictive distribution to two different regions of the feasible design space.}
	\label{fig:do_selection_example}
\end{figure}

Figure \ref{fig:do_selection_example} illustrates the benefits of preliminary CTO.
Suppose that, prior to undertaking CTO, we know only that the model outputs are positive.
Then $(0,0)$ is a natural choice as a target outcome.
However, the optimal region determined by the choice of $(0,0)$ in this case is somewhat arbitrary.
%
%
The point closest to $(0,0)$ is unique in the Pareto front solely in being nearest to the origin, and that choice of target outcome was itself driven merely by our ignorance of the feasible design space.
%
%
By contrast, suppose now that preliminary CTO has supplied us a rough estimate of the Pareto front, empowering us to choose a different target outcome -- for example, $(1.32,0.065)$ targets a point of diminishing returns in allowing $y_1$ to increase further in exchange for a reduced $y_2$.
%
%
%
%
Note also that when an emulator is used, preliminary CTO can use the same model observations as the subsequent CTO to train the emulator.
So preliminary CTO does not add to the budget of model runs, and is thus a computationally cheap supplement to CTO.

\section{Simulated Example}\label{example}
\begin{figure}
	\centering
	\includegraphics[scale=.85]{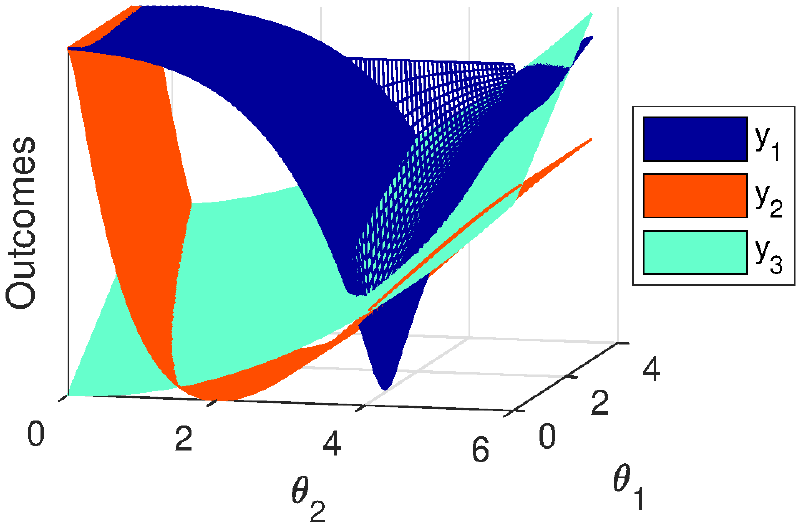}
	\caption{True outputs of the example model.}
	\label{fig:toy_sim_outputs}
\end{figure}

To illustrate our proposed procedure, consider the following problem of minimizing a function with trivariate output. 
Let $(x,\boldsymbol \theta)$ be the inputs, with scalar control input $x\in[1.95,2.05]$ and design variables $\boldsymbol \theta = (\theta_1,\theta_2)\in[0,3]\times[0,6]$.
We seek optimal settings for $\boldsymbol\theta$.
Model outputs are
$
y_1 = \left(\theta_1 \exp\left(-\left(\theta_1 + \lvert \theta_2-\frac{\pi x}2\rvert \right)\right)+1\right)^{-1}$, 
$
y_2 = \left(\theta_2^{x-1} \exp\left(-0.75 \theta_2\right) + 1 \right)^{-1}
$, and
$
y_3 = 15 + 2 \theta_1 + {\theta_2^2}/4.
$
We assume prior knowledge only that the outputs are positive.
Figure \ref{fig:toy_sim_outputs} displays the (normalized) outputs as functions of $\theta_1$ and $\theta_2$ at $x = 2$.
Assuming an easily evaluated model (so that an emulator is not needed), we have
$
\mathbf z(x) = \eta(x,\boldsymbol \theta) + \boldsymbol\epsilon
$
for target outcome $\mathbf z$, so that  $\boldsymbol\eta = (y_1,y_2,y_3)^T$ is the output and $\epsilon_i\sim N(\mathbf 0,\sigma_i^2)$ is the target observation error for $i=1,2,3$.
%
%

\begin{figure}
	\centering
	\includegraphics[scale=.85]{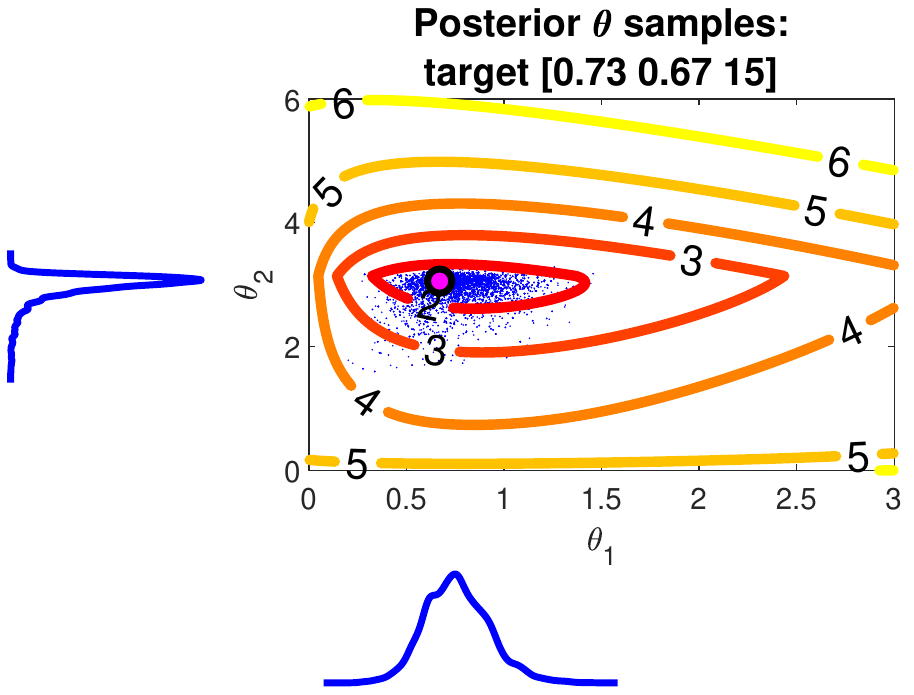}\\
	\vspace{1em}
	\includegraphics[scale=.85]{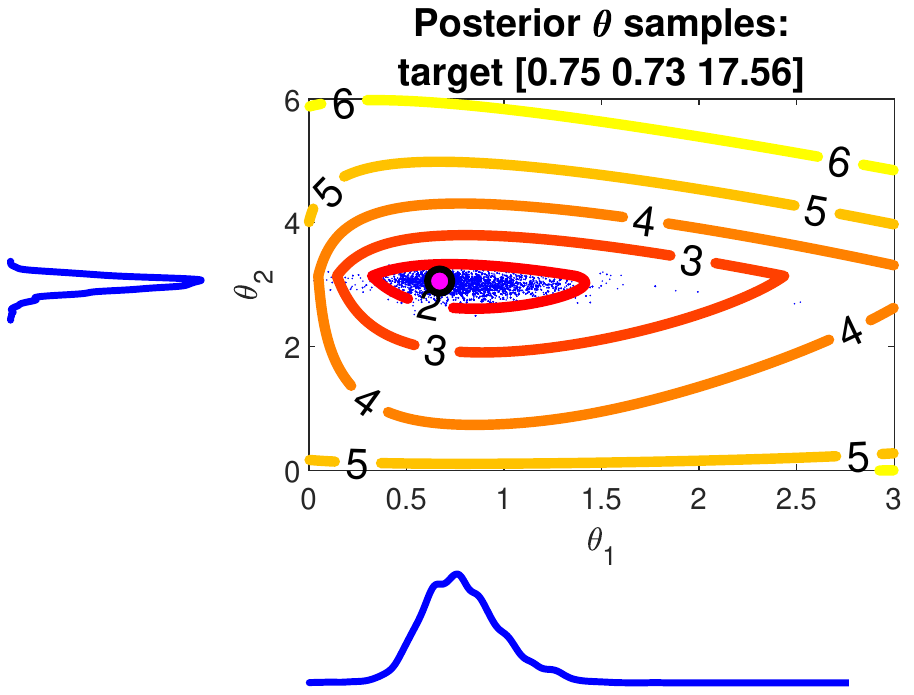}
	\caption{Posterior draws from CTO in the simulated example using the utopia point as a target (top) and using an updated target designed to lie one standard deviation from the Pareto front, in the direction of the utopia point (bottom). The contours show, for each point in the design space, the Euclidean distance of the model output at that point from the utopia point, averaged across the control input range $[1.95,2.05]$. The large dot shows the true optimum.}
	\label{fig:toy_sim_results}
\end{figure}

We initially set the target outcomes to $(0.7311, 0.6675, 15)$, the utopia point of the system, constant as a function of $x$. 
%
%
For comparison, we also performed CTO with target $(0.7506,\ 0.7302,\ 17.56)$, which lies close (one standard deviation away, under the uniform prior on $\theta$) to the feasible region, on the line connecting the original target utopia point to the nearest point in the feasible objective space.
The purpose of this comparison is to demonstrate the robustness of CTO to the distance separating the target outcome from the feasible objective space.
Thus a target outcome can be selected even when little is known about the location of the Pareto front.
Figure \ref{fig:toy_sim_results} shows the resulting posteriors. 
In the top plot, the target (the system's utopia point) is just over 1.6 units away from the objective space, where the each objective is standardized to have variance 1.
In the bottom plot, the Euclidean distance of the target from the objective space is 0.5.
Use of the utopia point produces a slightly longer lower tail for $\theta_2$, so that the posterior standard deviation of $\theta_2$ is increased from 0.1175 to 0.2126.
To minimize uncertainty regarding the optimal value of the design parameter, it is preferable to use a rough estimate of the Pareto front to select a target that is on or near the front, but when this is not possible, the resulting degradation of the results is typically minor.
The marginals in each case show substantial Bayesian learning compared to the prior (uniform) distribution of the design variables. 
CTO successfully maps the contours of the optimal region in each case, peaking near the true optimum. 

\section{Wind turbine material design application}\label{application}

In this section we use CTO to design a material for use in a wind turbine blade. 
The goal is to wed the typically separate tasks of material design and material selection for a specific application, designing a composite to optimize performance in a blade.
Our model uses \texttt{ANSYS} finite element analysis software \cite{ansys}. 
We assume the model accurately represents reality.

\subsection{Wind turbine blade design}

Two performance measures for blades are tip deflection and twist angle.
A goal of engineering design is to keep these measures and material cost low.
The blade is a composite of a given matrix and filler. 
%
%
The material properties (and thus blade performance and cost) depend on the thickness of the shear web in the blade and on the \emph{volume fraction}, or ratio of filler to matrix.
%
%
Temperature also affects the composite's properties and hence performance.
The model inputs are a triplet $(h,v,k)$, where $h$ is the temperature of the turbine (in kelvin), $v$ is the volume fraction, and $k$ is the thickness (in mm). 
The model output is a triplet $(d,r,c)$, where $d$ is tip deflection (in meters), $r$ is twist angle (in radians), and $c$ is cost per square meter (USD) of the material.
The turbine is deemed to operate over temperatures 230K-330K. 
%

\subsection{Emulation of finite element model}\label{emulator} 
The finite element simulator is too computationally expensive to be suitable for direct use in an MCMC routine. 
We employed a GP emulator in the manner of Ref.\ \cite{Williams2006}. 
For this purpose, we drew 500 (trivariate) observations from the finite element simulator according to a Latin hypercube sampling design \cite{McKay1979} based on plausible ranges for the three inputs as identified by subject matter experts: $[230\mathrm{K}, 330\mathrm{K}] \times [.2,.6]\times[10\mathrm{mm},25\mathrm{mm}]$.
We took the computer output to follow a GP with mean 0 and product power exponential covariance function as given in (\ref{eq:Hig_cov}).
The hyperparameters $\lambda_\eta,\boldsymbol \beta^\eta$ are estimated 
%
via maximum likelihood using only the finite element model output.
%
%
We used \texttt{fmincon()} \cite{MATLAB2017} 
to maximize (with $D=\boldsymbol\eta$) over the joint (four-dimensional) support of $\boldsymbol \beta^\eta,\lambda_\eta$.  
\begin{table}[h]
	\renewcommand{\arraystretch}{1.2}
	\begin{center}
		\begin{tabular}{|c|c|c|c|}
			\hline 
			& $d$ & $r$ & $c$ \\ 
			\hline 
			$\hat\rho^\eta_h$	&0.7239 & 0.7104  & 1 \\ 
			\hline 
			$\hat\rho^\eta_v$	&0.9788&  0.9723  & 0.9988 \\ 
			\hline 
			$\hat\rho^\eta_k$	& 0.9906 &0.9882  & 0.9986 \\ 
			\hline 
			$\lambda_\eta$	& 0.0177  & 0.0261 & 0.0009 \\ 
			\hline 
		\end{tabular} 
	\end{center}
	\caption{Covariance hyperparameter maximum likelihood estimates for each objective function. For each objective and each input $i$, $\rho_i^\eta = \exp(-\beta^\eta_i/4)$.}
	\label{table:mles}
\end{table}
The result, following the form of Equation \eqref{eq:Hig_cov} with $p=1$, $q=2$, and $(x,t_1,t_2)=(h,v,k)$ where for each objective $\rho^\eta_i = \exp(-\beta_i^\eta/4)$ for $i\in\{h,v,k\}$, appears in Table \ref{table:mles}.
%

\subsection{Design of the wind turbine blade system}\label{the_model}
\begin{figure}
	\centering
	\includegraphics[scale=0.85]{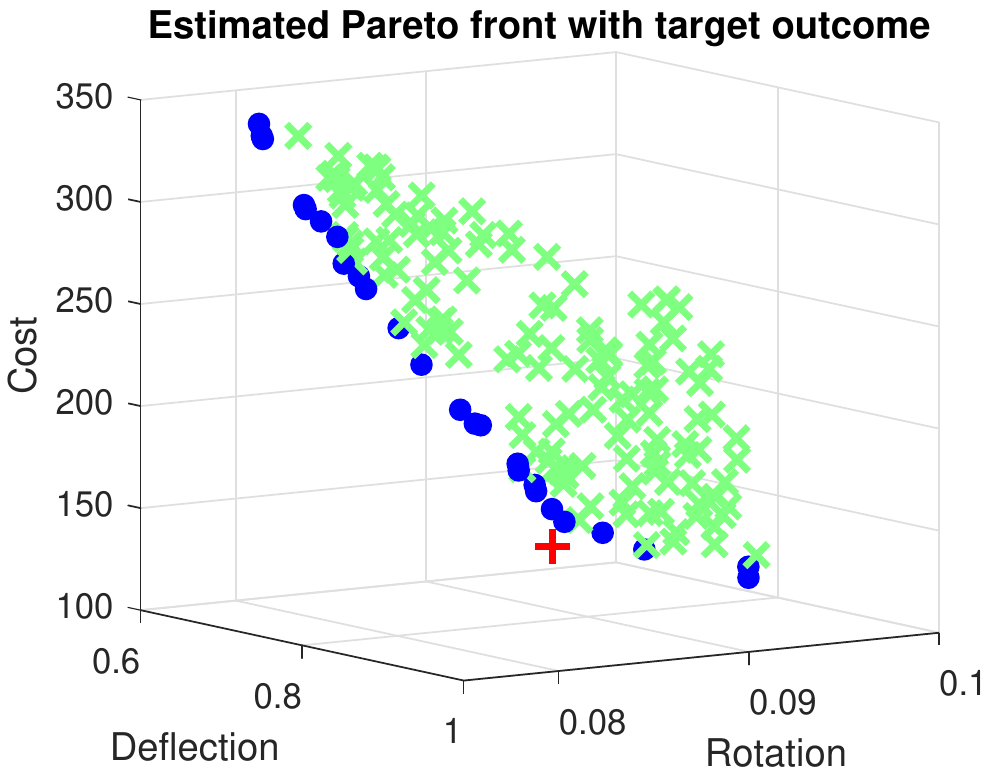}
	\caption{Each x is a dominated design drawn from the predictive distribution through preliminary CTO. The dots indicate the estimated Pareto front. The plus sign is the target selected as the performance objective in our proposed design approach.}
	\label{fig:elbow}
\end{figure}

\begin{figure}
	\centering
	\includegraphics[scale=0.85]{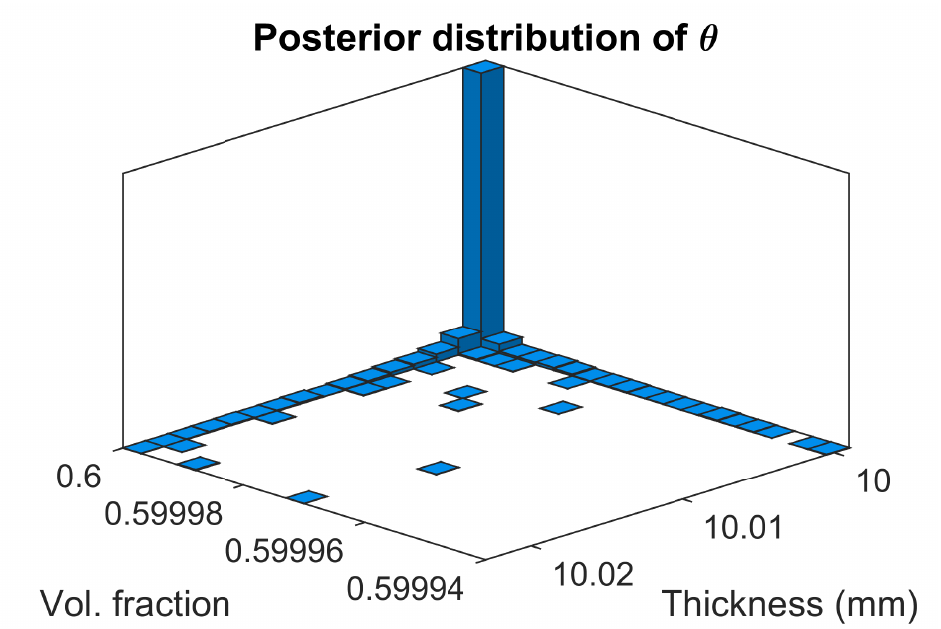}
	\caption{Histogram showing the posterior distribution from CTO in the wind turbine blade system. The prior is uniform over $[0.2,0.6]\times[10,25]$.}
	\label{fig:wt_marg_post}
\end{figure}

\begin{figure}[h]
	\centering
	\includegraphics[scale=0.85]{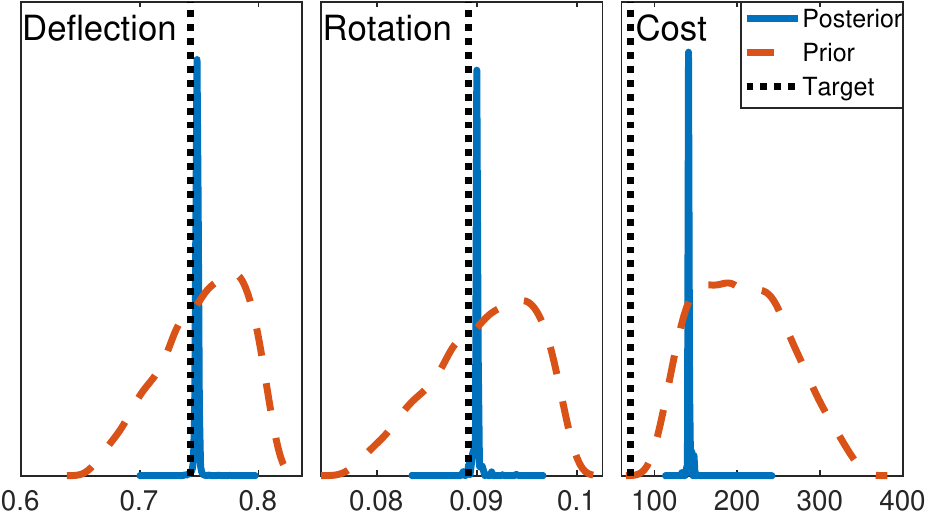}
	\caption{Approximate prior and posterior marginal predictive densities for each of the three outputs.}
	\label{fig:prior_post_pred_comp}
\end{figure}

All model inputs were rescaled to [0,1]. 
All model outputs were standardized so that each of the three responses has mean 0 and standard deviation 1.
The full joint posterior density of the design variables and discrepancy function hyperparameters is given in Equation \eqref{eq:full_dist}, using the MLEs given above.
Initial target outcomes were set to the estimated utopia point $(0.6551\mathrm{m},\ 0.0768\mathrm{rad},\ \$96.8)$ found by taking the minimum observed value of each objective from the 500 simulator observations. 
This target was set to be constant as a function of temperature, on an evenly-spaced grid of temperature values over the range [230K, 330K].
We carried out preliminary CTO with $\boldsymbol\sigma^2=5\cdot10^7\cdot(1,1,1)$ to estimate the Pareto front and locate a region of interest. 
For this purpose, 6,000 realizations were drawn via Metropolis-Hastings-within-Gibbs MCMC \cite{Metropolis1953, Hastings1970, Geman1984} in each of three chains (with random starts), of which the first 3,000 were discarded as burn-in. 
During the burn-in period, the covariances of the proposal distributions were periodically adjusted to be the sample covariance of the preceding draws scaled for an optimal acceptance rate of around $23\%$ for the multivariate $\boldsymbol \theta$ \cite{Roberts1997,Gelman2013}. 
%
%
%
Convergence of the three chains was verified visually and by the Gelman-Rubin statistic ($\approx1.01$ \cite{Gelman1992a}).
As expected for preliminary CTO, the posterior distribution of $\boldsymbol\theta$ was quite diffuse.
We used the GP emulator to predict the model output for each realization of $\boldsymbol \theta$.
Figure \ref{fig:elbow} displays the estimated Pareto front after filtering the posterior predictions to retain only non-dominated performance predictions.
Though the design space is three-dimensional, the Pareto front appears to be a roughly coplanar curve describing a trade-off between cost and deflection/twist.
A distinct point of maximum curvature appears in the Pareto front. 
This seems to be a point of diminishing returns in the trade-off between performance and cost, and thus we selected this point as the target for design.
To do so, we set the point $(\mathrm{deflection}=0.743\mathrm m,\ 
\mathrm{twist}=0.089\ \mathrm{rad},\ 
\mathrm{cost}=\$71.11)$
as the target outcome, constant as a function of temperature.
In the subsequent CTO, we employed the same MCMC approach as in the preliminary round, except we now estimate $\boldsymbol\sigma^2$, using a gamma(4,1/8) prior.
The covariances of the proposal distributions for each $\sigma^2_i$ were periodically adjusted to be the sample covariance of the preceding draws scaled for an optimal acceptance rate of around $44\%$ for the scalar $\sigma^2_i$ \cite{Roberts1997,Gelman2013}.
The posterior distribution of $\boldsymbol\theta$ appears in Figure \ref{fig:wt_marg_post}, almost as a point mass on $(0.6, 10\mathrm{mm})$.
Indeed, from the analysis discussed in Section \ref{removing_cal_pars}, we find that the ``elbow'' in the Pareto front is precisely the point at which volume fraction has reached its upper limit at $0.6$, with further gains possible only by raising thickness from its lower limit of $10$mm.
The contrast of the posterior distribution with the prior, which is uniform over $[0.2,0.6]\times[10,25]$, indicates that strong Bayesian learning has occurred.
%
The prior and posterior predictive distributions of the model outputs appear in Figure \ref{fig:prior_post_pred_comp}, where the prior predictive distributions are based on a uniform sampling of the model inputs.
The prior and posterior are not shown on the same scale, as the posterior is too sharp for both distributions to be visible on the same scale.
The mean output under the prior is $(0.753\mathrm m,0.091\text{ rads},\$206.58/\mathrm m^2)$, and under the posterior it is $(0.748\mathrm m,0.090\ \mathrm{rad},\$141.14/\mathrm m^2)$.
Though the mean performance outcomes are approximately the same under the posterior and the prior, mean cost per square meter and the uncertainty of the outcomes are dramatically lower.
If one prefers to prioritize gains in performance over cost, this can be accomplished by selecting target outcomes that reflect those priorities.
\subsection{Pareto front estimation with quantified uncertainties}\label{removing_cal_pars}

\begin{figure}[h]
	\centering
	\includegraphics[scale=.85]{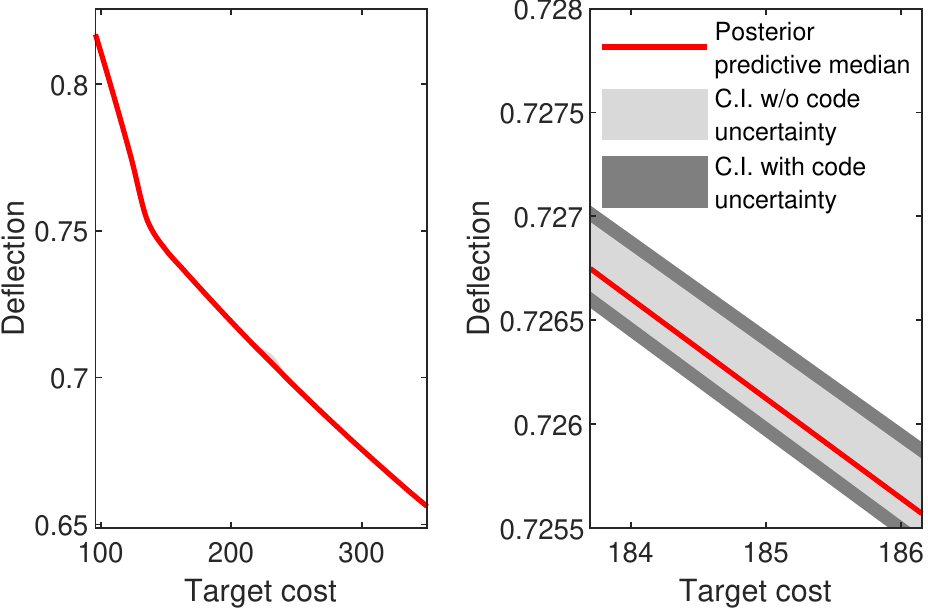}
	\caption{The estimated Pareto front of the wind turbine blade system with quantified uncertainties.}
	\label{fig:known_cost}
\end{figure}

%
When multiple design outputs are to be minimized, any point in the Pareto front is optimal relative to some set of priorities.
If those priorities have not been explicitly determined prior to the design process, then no particular outcome can be targeted.
%
%
For example, in a system where performance is monotonically increasing in cost, depending on one's tolerance for high cost, any point in the design space might be optimal.
%
%
In low-dimensional cases, CTO may be used to achieve a holistic picture of the Pareto front by optimizing to each target outcome on a grid.
To do this, where the model output is $b-$dimensional, one may draw a grid over the range of $b-1$ of the model outputs and perform CTO to minimize the remaining output at each point of the grid.
The $b-1$ outputs, at each grid point, are treated as known up to small observation error (e.g. $\sigma^2=0.01$ for objectives standardized to have mean 0 and variance 1).
Allowing some small observation error is necessary because the set of solutions having any particular exact value typically has probability 0.
The resulting estimate of the Pareto front differs from the filtering method employed in preliminary CTO in that it allows for quantifying the uncertainty associated with the Pareto front.
Our proposed procedure is illustrated here using the wind turbine blade application.
For simplicity, twist has been removed as a model output, leaving a system with 2-dimensional output of deflection and cost. 
The range of cost is known (via preliminary CTO) to be $[\$96,\$352]$.
A 20-point grid was drawn over this range of costs. 
%
%
For each point $c$ in the cost grid, we used the point $(0\mathrm m,\$c)$ as the target outcome for calibration (constant with respect to temperature).
The result is an estimate of the response surface with quantified uncertainty describing, for each point in the grid, the minimal achievable outcome for the output not included in the grid.
%
%
%
The result of applying this strategy to the wind turbine blade application is shown in Figure \ref{fig:known_cost}. 
The lefthand plot shows the estimated Pareto front for the system.
The associated uncertainty bands are too small to be distinguished in the lefthand plot, and so the righthand plot zooms in to show the bands for a portion of the Pareto front.
%
%
%

%
The use of CTO in the wind turbine design illustrates how preliminary CTO may be used, not merely to improve the identifiability of a pre-determined optimal region as in Section \ref{example}, but rather to identify a desirable region of the design space and select target outcomes that design to that region.
In the wind turbine case, selecting the utopia point as one's target determines the optimal region to be the high-cost region toward the upper-left of Figure \ref{fig:elbow}, since (on the standardized scale of model outputs) that region happens to be closest to the target.
If one has substantive goals that drive one to select that target, then one is well-served by optimizing to that high-cost region.
But if the utopia point is chosen arbitrarily, then the resulting optimal region is itself determined arbitrarily.
The estimate of the Pareto front provided by preliminary CTO allows us to identify regions of special interest, and to select target outcomes that that lead to clearly defined designs, as illustrated in Figure \ref{fig:elbow}.
The use of CTO in this case also demonstrates the value of obtaining a posterior distribution on the design variables, rather than just a point estimate.
For example, Figure \ref{fig:wt_marg_post} shows not just that a reasonable point estimate of the optimal $\boldsymbol\theta$ is at (0.6, 10mm)---respectively the upper and lower extrema of the supports for volume fraction and thickness---but also that if one input must deviate from its extremum it is better that the other one remain.
This tells against the idea that a reduction in volume fraction should be compensated by raising thickness.
More generally, CTO delivers an indication of the range of $\boldsymbol\theta$ values that achieve results near the target, which is potentially useful when one's goal is to set tolerances (rather than a specific value) for $\boldsymbol\theta$.
Finally, the use of CTO in the wind turbine case illustrates how the method can deliver ``Pareto bands'', providing not merely an estimate of the Pareto front (as in preliminary CTO) but also uncertainty associated with that estimate.
Such an estimate can be of use to decision-makers when deciding on performance goals subject to budgetary constraints.

\section{Conclusion} \label{conclusion}

We have described how the 
computer model calibration framework of  \cite{Kennedy2001} can be adapted for engineering design. 
Calibration to target outcomes undertakes design by ``calibrating'' a model not to field observations, but rather to artificial data representing performance and cost targets. 
The procedure optionally includes a computationally cheap preliminary step that provides a rough estimate of the Pareto front, which may be used to select target outcomes that promote strong Bayesian learning.
The resulting posterior predictive distribution approximates the target outcomes, so that the posterior distribution of $\boldsymbol\theta$ constitutes a distribution on optimal design settings.
Repeated applications of this methodology could allow one to construct a thorough estimate of the Pareto front of the system with quantified uncertainties by selecting target outcomes that explore different portions of the Pareto front.
Unlike other methods of Bayesian optimization (a review of which is provided by \cite{Shahriari2016}), CTO does not require the ability to evaluate model output adaptively.
Instead, it can rely on a batch of observations gathered prior to (and independently of) the design process.
We described the implementation of this approach in an MCMC routine along with considerations to accommodate computational instability.
The use of this methodology is illustrated in the case of material design for a wind turbine blade. 
%
%
By expropriating established tools of model calibration, CTO offers a method of optimization which is sensitive to, and quantifies, all sources of uncertainty.
The methodology as described here treats the computer model as universally valid over the domain of the design variables. 
Future work in this area will include the use of a discrepancy term capturing model bias.
%
%
Other possible extensions of our proposed methodology include its application to so-called ``state-aware calibration'' \cite{Atamturktur2015,Stevens2018,Brown2016}, which would allow the optimal region of the design variables to vary as a function of the control inputs.


%
%
%
%
%
%

\bibliographystyle{Chicago}

\bibliography{lit_review}
\end{document}